\documentclass[aps,prl,superscriptaddress,showkeys,floatfix,twocolumn]{revtex4-2}
\usepackage{graphicx}
\usepackage{amsmath}
\usepackage[colorlinks=true, allcolors=blue]{hyperref}

\newcommand{\lbr}{\langle}
\newcommand{\rbr}{\rangle}

\begin{document}

\title{Independent nuclear charge radius determinations: \\
bound-electron $g$ factor versus muonic atom spectroscopy}

\author{F. Hei\ss e}
\affiliation{Max-Planck-Institut für Kernphysik, Saupfercheckweg 1, 69117 Heidelberg, Germany}

\author{I.~A.~Valuev}
\affiliation{Max-Planck-Institut für Kernphysik, Saupfercheckweg 1, 69117 Heidelberg, Germany}

\author{J. Solanki}
\affiliation{Max-Planck-Institut für Kernphysik, Saupfercheckweg 1, 69117 Heidelberg, Germany}

\author{V. A. Yerokhin}
\affiliation{Max-Planck-Institut für Kernphysik, Saupfercheckweg 1, 69117 Heidelberg, Germany}

\author{B.~Ohayon}
\affiliation{The Helen Diller Quantum Center, Department of Physics, Technion-Israel Institute of Technology, Haifa, 3200003, Israel}

\author{N.~S.~Oreshkina}
\affiliation{Max-Planck-Institut für Kernphysik, Saupfercheckweg 1, 69117 Heidelberg, Germany}

\begin{abstract}
We present an alternative method to determine absolute nuclear charge radii competitively via the bound electron $g$ factor of H-like ions. The method is demonstrated for $^{118}$Sn by extracting its radius with a precision of $0.1\,\%$. This is the second most precise radius ever determined from an electronic system. We benchmark our result against a state-of-the-art value derived from the spectrum of muonic $^{118}\text{Sn}$. This comparison reveals a $3.2\,\sigma$ tension, prompting a careful reexamination of the calculations and calling for additional measurements. Crucially, after decades, we now have two independent methods to measure absolute charge radii with similar precision. 

\end{abstract}

\keywords{nuclear charge radii, bound-electron $g$ factor, muonic atoms
}

\maketitle

{\it Introduction --- } The root-mean-square (rms) nuclear charge radius is characteristic of every nucleus. It cannot be accurately predicted from first principles and influences all observables related to the electromagnetic interaction of the nucleus. Therefore, precise knowledge of this quantity is crucial for various fields, such as nuclear physics~\cite{Otten1989,Reinhard_2017,Yang2023,2025-isospin}, particle physics~\cite{Hardy2020,Gorchtein2024,2025-CKM}, precise tests of the Standard Model (SM), including bound state QED at high fields~\cite{Beiersdorfer2005,Paul2021,Akulov2025}, as well as the search for physics beyond the SM~\cite{Delaunay2017,2026-NP}.

Currently, absolute charge radii are primarily determined using two methods: elastic electron scattering and muonic-atom spectroscopy~\cite{Angeli2013Table,angeli2026}. Importantly, the extraction based solely on electron scattering cannot achieve competitive precision due to uncontrolled systematic uncertainties~\cite{FRICKE1995177, 2026-9Be, 2026-Cl, angeli2026, 2025-Def}. This often results in radii being extracted using only one method, without independent validation of the corresponding theories or experiments. 

The importance of an independent determination of charge radii was illustrated by high-precision comparisons between the proton radius extracted from muonic-atom spectroscopy and values inferred from electronic systems, which gave rise to the so-called proton radius puzzle~\cite{Pohl2010,Antognini_2013}. Due to the comparable uncertainties achieved by these approaches, their results can be directly confronted, instigating substantial progress in the field, including dedicated experimental initiatives~\cite{Bezginov2019, Maisenbacher2026} and refinements of the underlying theoretical framework~\cite{Antognini:2022ps,Pachucki_RevModPhys.96.015001}.

So far, state-of-the-art methods enable precise absolute radius comparisons between electronic and muonic atoms only for elements with $Z \leq 2$ and $Z \geq 70$ via the analysis of transition energies of low lying states. The respective data are only available for four nuclei of the more than $3000$ known nuclei: the proton, deuteron~\cite{Parthey2010, Pohl2016, Pachucki2018}, and, more recently, $^{208}\text{Pb}$ and $^{209}\text{Bi}$~\cite{Yerokhin_Pb_2026}. 

In this Letter, we address this gap by presenting a method based on electronic systems that enables the extraction of absolute radii with competitive precision for nuclei with $Z \gtrsim 40$. It resolves the finite nuclear size (fns) effect on the $g$ factor of H-like ions, which is currently measurable experimentally to 10 digits precision~\cite{2023Natur.622...53M}. 
Then we compare the result with a similarly precise value from muon atom spectroscopy.

\textit{Bound-electron $\bf \textit{g}$-factor} — We now extract the absolute nuclear charge radius of $^{118}\text{Sn}$ ($Z = 50$) by combining the high-precision experimental value of the bound-electron $g$ factor, the state-of-the-art QED theory, and nuclear charge shape parameters from electron-scattering experiments.  The bound-electron $g$ factor of $^{118}\text{Sn}^{49+}$ has been experimentally determined~\cite{2023Natur.622...53M} as
\begin{align}\label{eq1}
g_{\rm exp} = 1.910\,562\,059\,0\,(9)\,.
\end{align}

The current theoretical prediction for tin was obtained in Ref.~\cite{sikora:25} as
\begin{align}\label{eq2}
g_{\rm the} = 1.910\,561\,975\,(33)_{\rm the}(11)_{\rm rad}(19)_{\rm mod}\,,
\end{align}
where the first uncertainty is purely theoretical, the second arises from the tabulated value of the root-mean-square (rms) nuclear charge radius, $r_C = 4.6393\,(19)$~fm~\cite{Angeli2013Table}, and the third is due to the estimation of the nuclear charge model dependence. 

We extract the nuclear radius from the measured $g$-factor value~\eqref{eq1} in a model-independent way, using the method developed for energy levels in Ref.~\cite{Yerokhin_Pb_2026}. 
The method fixes the leading moments of the nuclear charge distribution, rendering the fns correction effectively independent of the nuclear model.
While Ref.~\cite{Yerokhin_Pb_2026} focused on fns corrections to energy levels, we extend the approach to the bound-electron $g$ factor; see End Matter for details.

The nuclear-shape effects are described by the following parameters
\begin{align}
V_{24} = \frac{\lbr r^2 \rbr^{1/2}}{\lbr r^4 \rbr^{1/4}}\,, \ \ \ 
V_{26} = \frac{\lbr r^2 \rbr^{1/2}}{\lbr r^6 \rbr^{1/6}}\,.
\end{align}
In this work, we extract $V_{24}$ and $V_{26}$ from electron-scattering data by integrating the charge distributions in Refs.~\cite{1974-Sick,1987-JDV} and extrapolating to the high-momentum-transfer limit. For $^{118}$Sn we obtain
\begin{align}
V_{24} = 0.9350\,(16)\,, \ \ \ \ \ V_{26} =  0.888\,(3)\,. \label{nuclear_moments}
\end{align}
The uncertainties were estimated as the difference between values inferred from measurements at the highest available momentum transfer and the corresponding extrapolated values.
The electron-scattering values of $V_{24}$ and $V_{26}$ could be cross-checked in the future by nuclear theory calculations~\cite{Miyagi2025_Rch4,2026-taming}.

The theoretical $g$-factor value given by Eq.~(\ref{eq2}) was obtained in Ref.~\cite{sikora:25} using the nuclear radius $r_{C,0} = 4.6393~\textrm{fm}$ and the standard Fermi model, corresponding to the following values of the $V$ coefficients:
$V_{24,0} = 0.93257\,,$ and $V_{26,0} = 0.88170\,.$ 
We therefore express the theoretical $g$-factor value for the fixed nuclear parameters as
\begin{align}\label{eq5}
g_{\rm the}\left(r_{C,0},V_{24,0}, V_{26,0}\right) = 1.910\,561\,975\,(33)\,.
\end{align}
The difference between the theoretical value above and the experimental result of Eq.~(\ref{eq1}) should arise from the shift of the fns correction caused by the nuclear parameters
\begin{align} \label{eq6}
\delta_{\rm fns}g = 
g_{\rm fns}\left(r_{C},V_{24}, V_{26}\right) - 
g_{\rm fns}\left(r_{C,0},V_{24,0}, V_{26,0}\right) \,.
\end{align}
By requiring that $\delta_{\rm fns}g = g_{\rm exp} - g_{\rm the}\left(r_{C,0},V_{24,0}, V_{26,0}\right)$, we determine the nuclear rms radius $r_C$. 

The fns correction to the bound-electron $g$ factor is represented as \cite{Yerokhin_2013}
\begin{align}
         g_{\rm fns} = g_N \left(1 + \frac{\alpha}{\pi}\,G_{\rm QED}\right)\,,
\end{align}
where $g_N$ is the fns correction to the Dirac value of the bound-electron $g$ factor and the function $G_{\rm QED}$ represents the corresponding QED correction. For the QED correction, we take the value $G_{\rm QED}(Z = 50) = -1.821\,(1)$, as calculated in Ref.~\cite{Yerokhin_2013}, and assume that it is independent of the nuclear parameters. This approximation is justified by the small magnitude of the correction coupled with the minor shift in the nuclear radius.

To determine $r_C$, it is convenient to first compute $\delta_{\rm fns}g$ for many different sets of nuclear-model parameters in a vicinity of $\left(r_{C,0},V_{24,0}, V_{26,0}\right)$ and then fit the numerical results to the form \cite{Yerokhin_Pb_2026}
\begin{equation}\label{gfs2}
        \delta_{\rm fns} g = C_{2\gamma}\, 
        \delta \lbr r^2\rbr^\gamma + C_4\, \delta \lbr r^4\rbr + D_4\, \delta \lbr r^2\rbr ^2 + C_6\, \delta \lbr r^6\rbr \,,
\end{equation}
where $\delta \lbr r^n\rbr = \lbr r^n\rbr - \lbr r^n_0\rbr$, $\gamma = \sqrt{1-(Z\alpha)^2}$, and $C_i$ and $D_{4}$ are fitting coefficients.
Once the fitting coefficients are obtained, determining $r_C$ is straightforward. 
Our practical calculations were performed using the three-parameter Fermi model as given by Eq.~(A4) of Ref.~\cite{Yerokhin_Pb_2026}, but the method ensures that the results are model-independent.

Our fitting yields the following results for the coefficients of Eq.~(\ref{gfs2}) (with all distances expressed in femtometers):
$ C_{2\gamma} = 8.2923 \times 10^{-7}$,
        $C_4 = -4.0627 \times 10^{-10}$,
        $D_4 = 3.6736 \times 10^{-10}$, and
        $C_6 = 1.023 \times 10^{-12}$.

Combining these fitted coefficients with the nuclear-moments ratios from Eq.~(\ref{nuclear_moments}),
we obtain the following result for the rms charge radius of $^{118}\text{Sn}$: 
\begin{equation}\label{rms}
    r_C = 4.6536\,(57)_{\rm the}\,(2)_{\rm exp}\,(3)_{\rm mod}~\text{fm}\,. 
\end{equation}
Here, the first uncertainty comes from the bound-state QED description of the bound-electron $g$ factor, the second uncertainty originates from the experimental value of the $g$ factor, and the third is the nuclear-model uncertainty due to errors of $V_{24}$ and $V_{26}$. If the theoretical $g$-factor prediction $g_{\rm the}$ or the values of the $V$ coefficients are updated, the rms radius can be readily reevaluated using Eq.~(\ref{gfs2}) and the fitting coefficients given in the text.

To the best of our knowledge, this is the first precise extraction of an absolute nuclear charge radius from the bound-electron $g$ factor. It is also the second most accurate absolute radius determined solely from electronic systems (after the deuteron charge radius obtained as the sum of the proton radius \cite{Maisenbacher2026} and the proton-to-deuteron isotope difference~\cite{Parthey2010}), surpassing the relative precision of the proton radius~\cite{Maisenbacher2026}. Additionally, this is the first nucleus heavier than the deuteron for which it is possible to compare absolute nuclear charge radii with a relative accuracy of 0.1\,\%. 

\textit{Muonic atom spectroscopy} --- We also analyze muonic atom spectroscopy data to benchmark our result and enable comparison between both methods. The spectroscopy of muonic atoms is a well-established technique for determining nuclear charge radii~\cite{BorieRinker1982, Nortershauser2020, Angeli2013Table, Fricke2004} due to the high sensitivity of muonic systems to nuclear parameters. 

Unlike electronic atoms, whose energies are predominantly affected by nuclear size, the energy levels of muonic atoms depend strongly on both size and shape. Each transition has its own dependence, so that when enough lines have been measured with sufficient precision, both size and shape can be determined independently of information from electronic systems. Here we use this approach to facilitate a robust benchmark for the $g$-factor determination.

A key distinction between the case of muonic Sn and a lighter muonic system, or electronic $g$-factor is that the nuclear size dependence is no longer factorizable~\cite{2025_muonic_Cl}. Consequently, nuclear effects must be included  non-perturbatively, already at the level of the Dirac equation. This necessitates a full ab initio calculation of all relevant contributions across a range of nuclear parameters, followed by a global fit to the experimental data. For the current evaluation of the rms radius from muonic data, we analyzed experimental data taken from~\cite{Fricke2004, ENGFER1974509}. The theoretical framework is detailed in our earlier works \cite{Pb_PhysRevLett.135.163002, beyer2025relativistic}; we summarize it briefly here. We employed two-parameter Fermi distribution (2pF) to model the nuclear density. First-order quantum electrodynamics corrections in $\alpha$ were included: the leading Uehling, Wichmann-Kroll, and self-energy corrections, as well as nuclear recoil, electron screening, and nuclear polarization effects. The inclusion of the Uehling potential within the Dirac Hamiltonian accounts for part of the second-order vacuum-polarization effects, the remaining part is described by Kallen-Sabry correction, which was also incorporated.  

Due to strong pairing correlations and the complex collective nature of low-lying nuclear states in $^{118}\text{Sn}$, achieving high-accuracy microscopic descriptions of these excitations is extremely challenging. Therefore, the nuclear-polarization corrections were calculated using the approach presented in~\cite{2025_muonic_Cl}, wherein the parameters for the low-lying excitations were taken directly from experimental data, while the giant resonances were accounted for collectively by means of energy-weighted sum rules.
As a result, the nuclear-polarization uncertainty dominates the theoretical error budget and is estimated to be $\approx 25\%$ of the effect, which is about twice as large as the microscopic-model uncertainties for low-lying muonic states in the case of semi-magic $^{120}\text{Sn}$~\cite{beyer2025relativistic}.

\begin{table}[tbp]
    \centering
    \begin{tabular}{l l l}
    \hline \hline
     Line  \qquad\qquad  &  Experiment  & Best-fit value \\ \hline
 $2p_{1/2}-1s_{1/2}$ & 3414.352(112)~\cite{Fricke2004} & 3414.08(56) \\ 
 $2p_{3/2}-1s_{1/2}$ & 3459.748(95)~\cite{Fricke2004} & 3459.90(56) \\
 $3d_{3/2}-2p_{1/2}$ & 1022.8(2)~\cite{ENGFER1974509} & 1022.60(1)\\
 $3d_{5/2}-2p_{3/2}$ & \phantom{1}982.6(2)~\cite{ENGFER1974509} & \phantom{1}982.75(1) \\
     \hline\hline
    \end{tabular}
    \caption{Experimental and the best-fit theoretical values for the transition energies of muonic $^{118}$Sn, in keV. The parasynthesis for the theory values stands for the total uncertainties and dominated by the uncertainty of NP contribution.}
    \label{tab:mu_sn118}
\end{table}

\begin{figure}[tbp]
    \centering
    \includegraphics[width=0.99\linewidth]{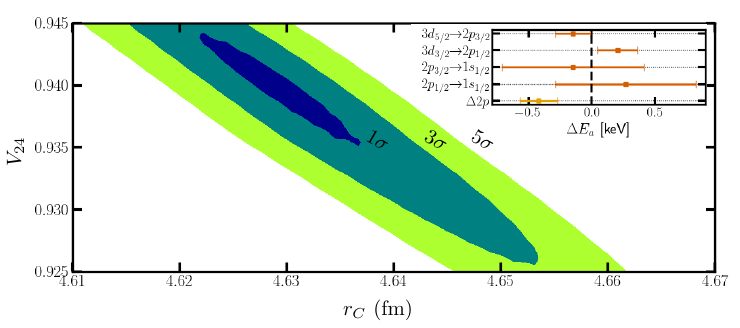}
    \caption{The root-mean-square radius $r_C$ and $V_{24}$ ratio extracted from the fit of the $\mu^{118}$Sn spectrum to a two-parameter Fermi  distribution. The colored regions show the 5, 3, and 1$\sigma$ confidence intervals. The insets show the residual transition energies with their 1$\sigma$ uncertainties (combined experimental and theoretical contributions). Measured transitions are shown in darker orange, while the derived fine-structure splittings are shown in lighter color. }
    \label{fig:muonic_rms_v24}
\end{figure}

To perform the fit, we computed transition energies on a grid for $a$ and $c$ parameters of 2pF. As previously observed in our studies, the resulting best-fit values of the 2pF parameters $a$ and $c$ are less precise than the rms value due to the strong correlations between them. The same trend is evident here, with the nuclear radius value 
\begin{equation}
r_C = 4.6291\,(51)\,\text{fm}, 
\end{equation}
with 2pF parameters of $c=5.59\,(7)$\,fm and $a=0.44\,(4)$\,fm. Moreover, the quality of the fit is $\chi^2/{\rm DoF} =4.2$ and the corresponding ratios of the nuclear moments are $V_{24} = 0.940\,(3)$ and $V_{26}=0.896\,(7)$. The used experimental data together with the best-fit theory predictions are shown in Tab.~\ref{tab:mu_sn118}. The overview of the results in terms of rms value and $V_{24}$ parameter is shown in Fig.~\ref{fig:muonic_rms_v24}. 

\begin{figure}[!tb]
    \centering
    \includegraphics[width=8cm]{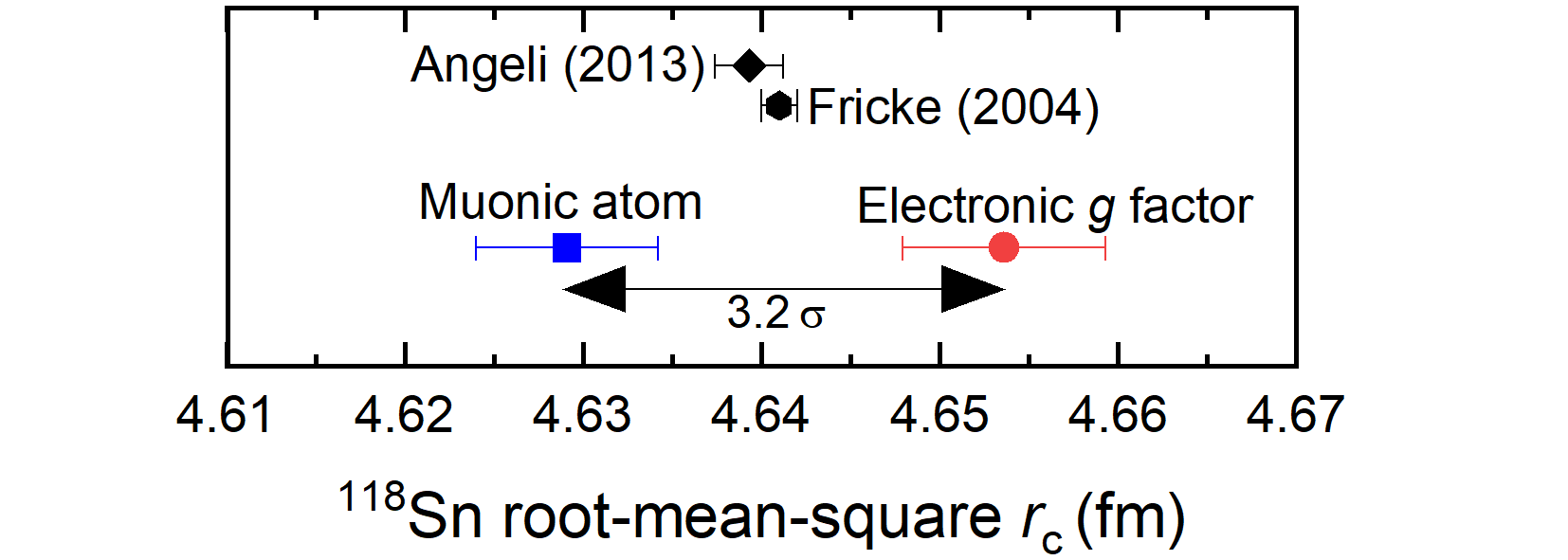}
    \caption{The rms nuclear charge radii of $^{118}\text{Sn}$ of the electronic and muonic based methods are in tension by combined $3.2\,\sigma$. The literature values of Angeli \textit{et al.}~\cite{Angeli2013Table} and Fricke \textit{et al.}~\cite{Fricke2004} are derived from a combined analysis of electron scattering, muonic spectroscopy and optical isotope shift measurements. An update of these compilations is currently underway~\cite{angeli2026}.}
    \label{fig:electronic_vs_muonic_rms}
\end{figure}

\textit{Discussion} --- Our two independent results for the absolute nuclear charge radius of $^{118}$Sn are:
\begin{align*}
    r_C\text{[$g$ factor]} &= 4.6536\,(57) \text{ fm,} \\
    r_C\text{[muonic]} &= 4.6291\,(51) \text{ fm}\, .
\end{align*}
They have similar accuracy and a $3.2\,\sigma$ tension between them, as illustrated in Fig.~\ref{fig:electronic_vs_muonic_rms}. 
In the following, we analyze potential factors that may contribute to this deviation.

The most obvious concern is the measured muonic spectrum, combined from two experimental data sets \cite{Fricke2004, ENGFER1974509}. Additionally, a detailed assessment reveals a discrepancy of approximately $3\,\sigma$ in the $\Delta 2 p$ fine-structure interval. This indicates an experimental issue, as the corresponding interval in $\mu^{120}\text{Sn}$ is consistent with theoretical predictions, with an associated $\chi^2/{\rm DoF} = 0.9$~\cite{beyer2025relativistic}. However, rescaling all quoted uncertainties for $\mu^{118}\text{Sn}$ so that the goodness-of-fit is normalized to unity would reduce the discrepancy between the muonic and $g$-factor determinations to $2.1\,\sigma$, thereby providing only partial mitigation of the tension.

The next possible contribution to the tension could be incomplete knowledge of nuclear structure, which is crucial for evaluating the nuclear-polarization correction. All other theoretical contributions have been calculated using the well-established QED approaches, which have been extensively tested in highly-charged ions and muonic atoms, and have generally yielded consistent results, including in applications to muonic transitions in systems very similar to $\mu^{118}\text{Sn}$~\cite{Pb_PhysRevLett.135.163002, beyer2025relativistic}. Nuclear shape effects, neglected in 2pF model, could potentially contribute to the discrepancy. However, in practice we do not expect them to contribute significantly, especially since the values of $V_{24, 26}$ from muonic atoms and from electron scattering are in reasonable agreement. 

With respect to the $g$-factor measurement, the reported experimental uncertainty is sufficiently small that reconciling the observed deviation would require an underestimation by approximately two orders of magnitude, a scenario that appears highly improbable.
The same applies to the effect of $V_{24}$ and $V_{26}$ on the extraction.

The more likely source of the tension is an underestimated theory contribution to the bound-electron $g$ factor calculations, which might apply to the recently developed two-loop all order calculation~\cite{sikora:25}.

Eventually, since both cases test the strongest electric fields (real electron-positron pairs are produced during the $2p-1s$ transition in $\mu^{118}$Sn~\cite{Andreev2026}), deviations from QED predictions are conceivable; therefore, unknown contributions cannot be completely excluded~\cite{Mohr_1998,Beier00}.

\textit{Outlook} --- The reported tension between muonic and $g$-factor values or the rms charge radius of $^{118}$Sn has prompted further experimental checks. First, new muonic atom spectroscopy experiments with increased precision for $\mu^{118}$Sn would be beneficial. At the same time, since the recently published results for $\mu^{120}$Sn \cite{beyer2025relativistic} are significantly more accurate, a corresponding measurement of the bound-electron $g$ factor for $^{120}\text{Sn}^{49+}$ is very valuable to clarify this discrepancy.

It is important to emphasize that the uncertainty of our $g$ factor radius determination is entirely dominated by theoretical uncertainties. Calculations to reduce these uncertainties are currently underway~\cite{volkov:priv}.
Furthermore, our analysis demonstrates that the dependence of the nuclear charge radius on nuclear-polarization effects and the ratio of the leading moments of the nuclear charge distributions for the electron $g$ factor are suppressed by an order of magnitude compared to the muonic atom spectroscopy result~\cite{Choudhury2026NuclearPolarization}. Therefore, results based on the electron $g$ factor have the potential to provide significantly less model-dependent and improved nuclear charge radii; however, this comes at the expense of considering the nuclear-moments ratio from another source such as electron scattering or theory~\cite{Miyagi2025_Rch4,2026-taming}.

A significant advantage of determining nuclear radii from the bound-electron $g$ factor, compared to their determination from transition energies~\cite{Yerokhin_Pb_2026}, is that the $g$ factor of an $s$ state of any ion with a spinless nucleus is close to the free-electron value, $g \approx 2$, allowing for precise measurements across a wide range of elements using essentially similar experimental setups. Importantly, measurements with a precision of $10^{-10}$ have already been demonstrated for several ions in the range $Z = 2-50$~\cite{Schneider2022,Sturm_nature_2014,Heisse2023,Sturm2013}, with extensions to higher-$Z$ ions and elements with non-zero nuclear spin anticipated in the near future~\cite{sturm_alphatrap_2019,Rausch2026,Yerokhin_2011}.

Finally, we emphasize that both values may be consistent with the available evidence, and that there exists a potentially significant distinction between the short-distance dynamics involving muons and those involving electrons. Consequently, the possibility of physics beyond the SM cannot be ruled out~\cite{Moretti2026,Quint2026,moretti2026b,PhysRevA.92.012123,PhysRevLett.107.011803,PhysRevLett.117.101801,Crivellin:2024zui}.

\begin{acknowledgments} 
\textit{Acknowledgments} --- This work is supported by the Collaborative Research Centre 1225 funded by Deutsche Forschungsgemeinschaft (DFG, German Research Foundation) Project No. 273811115–SFB 1225 ISOQUANT. This work comprises parts of the PhD thesis work of J. S. to be submitted to Heidelberg University, Germany.
\end{acknowledgments}

\textit{Data availability} --- The data that support the findings of this article are not publicly available. The data are available from the authors upon reasonable request.

\bibliography{refs}

\section*{End Matter}

\textit{Bound-electron $\bf\textit{g}$-factor} --- It was demonstrated in Ref.~\cite{Yerokhin_Pb_2026} that the model dependence of finite-nuclear-size (fns) corrections to the energy levels of electronic atoms can be systematically eliminated by constraining an increasing number of moments of the nuclear charge distribution. It was also found that if a model of the nuclear charge distribution reproduces the first three moments of the nuclear charge distribution, the residual model dependence becomes completely negligible at the level of present-day accuracy requirements.

Table~\ref{tab:fns_sn118} extends the analysis of Ref.~\cite{Yerokhin_Pb_2026}, presented in Table~I of that work, to the bound-electron $g$ factor. Based on the data presented in Table~\ref{tab:fns_sn118}, we conclude that the situation for the bound-electron $g$ factor is completely analogous to that for the energy levels and that the model dependence of the theoretical $g$-factor values can be eliminated in the same way.

\begin{table*}
\centering
\caption{Finite nuclear size correction to the energy ($E_{\rm fns}$) and to the bound electron $g$-factor ($g_{\rm fns}$) for the
$1s$ state of H-like tin ion $^{118}\mathrm{Sn}^{49+}$, for different models of the
nuclear charge distributions. $n$ denotes the number of
constrained nuclear-charge moments. 
$E_{\rm fns}$ denotes the fns energy corrections, $g_{\rm fns}$ denotes
the fns corrections to the bound electron $g$-factor, 
and ``Diff.'' denotes
the deviation from the Fermi model with the same $n$.\\}
\label{tab:fns_sn118}

\small
\setlength{\tabcolsep}{4pt}
\begin{tabular}{c l c c c c c c c c}
\toprule
$n$ & Model &  $r_C$ [fm] &
$V_{24}$ & $V_{26}$ & $V_{28}$ &
$E_{\rm fns}$ [eV] & Diff. & $ g_{\rm fns}$ ($\times 10^{6}$) & Diff. \\
\hline

1
& 1pF &  4.6393 & 0.9326 & 0.8817 & 0.8388 &
1.94568 & & 14.4900 & \\
& 1pH &  4.6393 & 0.9573 & 0.9302 & 0.9112 &
1.94810 & $\ \ 0.12\%$ & 14.5084 & $\ \ 0.13\%$ \\
& 1pG &  4.6393 & 0.8801 & 0.7974 & 0.7356 &
1.93923 & $-0.33\%$ & 14.4407 & $-0.34\%$ \\
& 1pE &  4.6393 & 0.7953 & 0.6640 & 0.5718 &
1.92568 & $-1.03\%$ & 14.3369 & $-1.06\%$ \\

2
& 2pF &  4.6393 & 0.935 & 0.8864 & 0.8458 &
1.94593 & & 14.4919 & \\
& 2pH &  4.6393 &
0.935 & 0.8962 & 0.8706 & 1.94573 & $-0.010\%$ & 14.4904 & $-0.010\%$ \\
& 2pG &  4.6393 &
0.935 & 0.8953 & 0.8655 & 1.94573 & $-0.010\%$ & 14.4904 & $-0.010\%$ \\
& 2pE & 4.6393 &
0.935 & 0.8632 & 0.7686 & 1.94627 & $\ \ 0.018\%$ & 14.4945 & $-0.018\%$ \\

3
& 3pF &  4.6393 &
0.935 & 0.888 & 0.8501 & 1.94590 & & 14.4917 & \\
& 3pH &  4.6393 &
0.935 & 0.888 & 0.8516 & 1.94590 & $\ \ 0.0003\%$ & 14.4917 & $\ \ 0.0003\%$ \\
& 3pG &  4.6393 &
0.935 & 0.888 & 0.8504 & 1.94590 & $\ \ 0.0002\%$ & 14.4917 & $\ \ 0.0002\%$ \\
& 3pE &  4.6393 &
0.935 & 0.888 & 0.8472 & 1.94589 & $-0.0003\%$ & 14.4916 & $-0.0003\%$ \\

\hline\hline
\end{tabular}
\end{table*}

\end{document}